\documentclass[12pt]{iopart}
\usepackage{graphicx}
\usepackage{color}
\usepackage{caption}
\usepackage{soul}
\usepackage{siunitx}
\usepackage{amssymb,latexsym} 

\begin{document}

\title{Coherent perfect absorption in resonant materials}

\author{S. Shabahang, Ali. K. Jahromi, L. N. Pye, J. D. Perlstein, M. L. Villinger, and Ayman F. Abouraddy}

\address{CREOL, The College of Optics \& Photonics, University of Central Florida, Orlando, Florida 32816, USA}
\ead{jahromi@creol.ucf.edu}
\vspace{10pt}
\begin{indented}
\item[]September 2017
\end{indented}

\begin{abstract}
Coherent perfect absorption (CPA) is an interferometric effect that guarantees full absorption in a lossy layer independently of its intrinsic losses. To date, it has been observed only at a single wavelength or narrow bandwidths, whereupon wavelength-dependent absorption can be ignored. Here we produce CPA over a bandwidth of $\sim$ 60~nm in a 2-$\mu$m-thick polymer film with a low-doping concentration of an organic laser dye. A planar cavity is designed with a spectral `dip' to accommodate the dye resonant linewidth, and CPA is thus achieved even at its absorption edges. This approach allows realizing strong absorption in laser dyes -- and resonant materials in general -- independently of the intrinsic absorption levels, with a flat spectral profile and without suffering absorption quenching due to high doping levels.
\end{abstract}

%
\vspace{2pc}

Maximizing optical absorption is at the heart of enhancing the performance of a variety of photonic devices \cite{Baranov17NMR}, such as photovoltaics \cite{Yu10PNAS,Villinger19unpubl}, photodetectors \cite{Muller78IEEE,Unlu95JAP}, and laser gain media \cite{Kozlovsky92JQE}, and can even help introduce new functionalities \cite{Fang2015Light,Papaioannou16APLPhotonics}. This goal can be realized via coherent perfect absorption (CPA) – an interferometric effect that guarantees complete absorption in a material independently of its intrinsic level \cite{Chong10PRL,Wan11Science,Zhang12Light,Villinger15OL}. This powerful capability has prompted extensive efforts for realizing CPA in numerous settings including plasmonics and metasurfaces \cite{Zhang12Light,Yoon12PRL,Jang14PRB,Roger15NatCommun,Jung15OE,Pirruccio16PRL}, semiconductor \cite{Ramakrishnan13OE, Xu18MSEB} and metal \cite{Pohlack92SPIE,Heger04SPIE} films, 2D materials \cite{Pirruccio13ACSNano,Rao14OL,Huang16ACSNano}, optical fibers \cite{Jahromi18IEEE}, planar structures \cite{Schmidt20APL,Kats12APL,Furchi12NanoLtt,Villinger15OL,Kakenov16ACSPhotonics,Pye17OL}, and on-chip systems \cite{Rothenberg16OL,Zhao16PRL,Wong16NatPhoton}. However, all these realizations were made over narrow bandwidths or even at a single wavelength. \textit{Broadband} CPA is a more demanding task because it must contend with wavelength-dependent absorption, which is most prominent in materials featuring resonant absorption, such as quantum dots, perovskites, and organic laser dyes -- which are yet to be investigated in this context.

\begin{figure}[b!]
\centering\includegraphics[width=8.6cm]{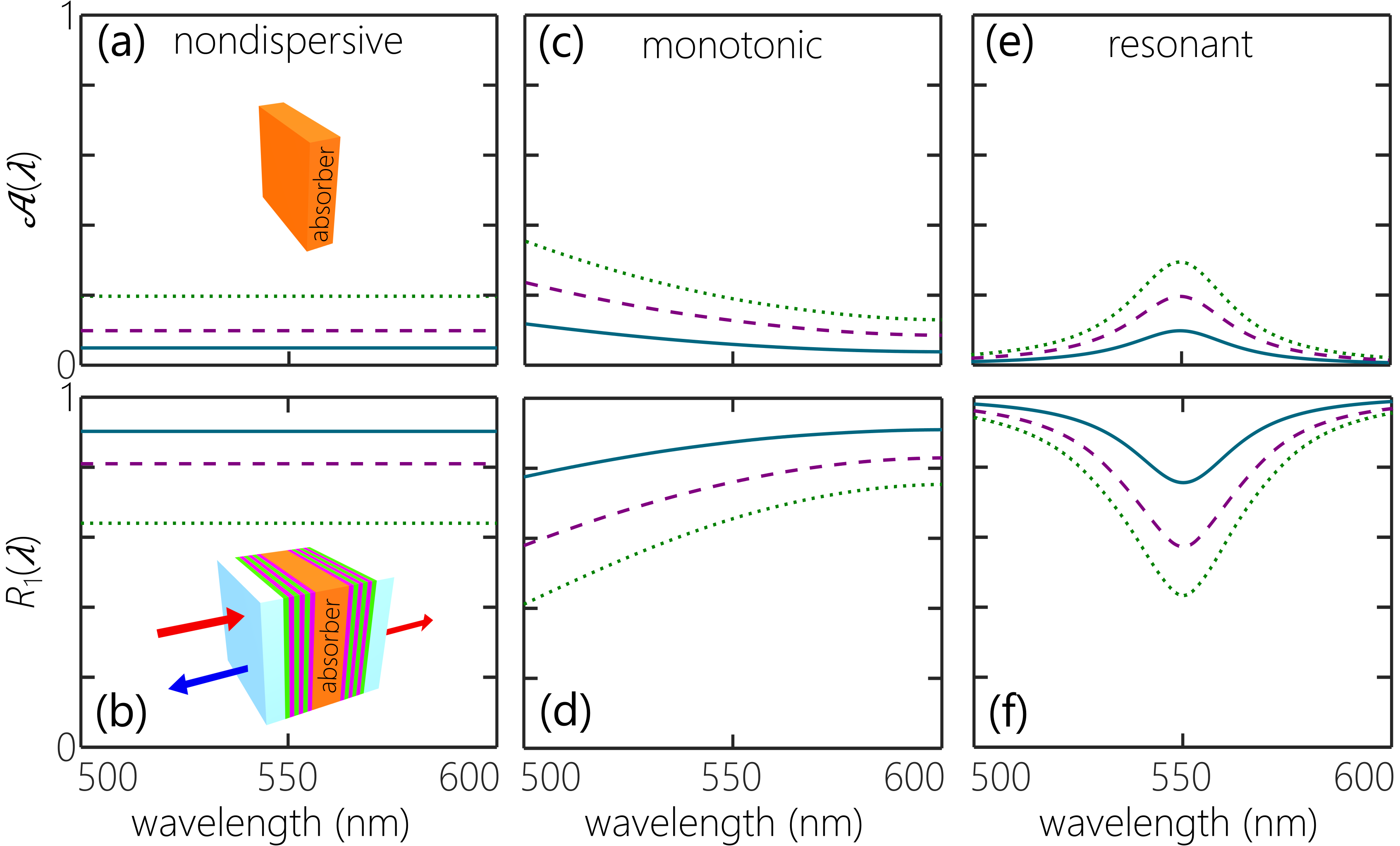}
\caption{\small{(a) Flat single-pass absorption $\mathcal{A}(\lambda)$ and (b) the reflectivity $R_{1}(\lambda)\!=\!(1-\mathcal{A})^{2}$ required to achieve CPA ($\mathcal{A}_{\mathrm{L}}\!=\!1$) in the cavity shown in the inset. Dotted, dashed, and solid curves correspond to three different values for $\mathcal{A}$. (c,d) Same as (a,b) for a layer with monotonically decreasing $\mathcal{A}(\lambda)$. (e,f) Same as (a,b) for a layer with resonant absorption.}}
\label{Fig:Concept}
\end{figure}

Here we realize CPA in weakly absorbing 2-$\mathrm{\mu m}$-thick spin-coated polymer films impregnated with laser dyes: rhodamine 101 dye and DCM (4-(Dicyanomethylene)-2-methyl-6-(4-dimethylaminostyryl)-4H-pyran) dye, which exhibit resonant absorption in the visible. This is achieved by embedding the doped polymer film in a planar cavity that incorporates aperiodic multilayer dielectric mirrors whose reflectivity is specially designed to inversely track the dispersive absorption of the dyes. By introducing an appropriate spectral `dip' in the mirror reflectivity, spectrally flat coherently enhanced absorption is produced at cavity resonances spanning a bandwidth ($\sim60$~nm) larger than the dye linewidth. Rather than optimize the interference of counterpropagating beams within the absorbing structure \cite{Wan11Science,Schmidt20APL,Zhang12Light}, we critically couple a single beam to the cavity \cite{Pye17OL}, which is a more appropriate configuration for applications in photovoltaics and photodetection. This strategy can help circumvent the challenge of concentration quenching in highly doped thin films by guaranteeing complete optical absorption independently of the doping level. Indeed, we observe an absorption enhancement by a factor of $\sim12-25$ in the weakly absorbing, low-dye-concentration 2-$\mathrm{\mu m}$-thick polymer films. Furthermore, we realized this effect in symmetric and asymmetric cavities, demonstrating the superiority of the latter when provided with a back-reflector. Finally, we show that adding a spacer to reduce the cavity free spectral range (FSR) does \textit{not} impact the achievable absorption.


\begin{figure}[t!]
\centering\includegraphics[width=8.6cm]{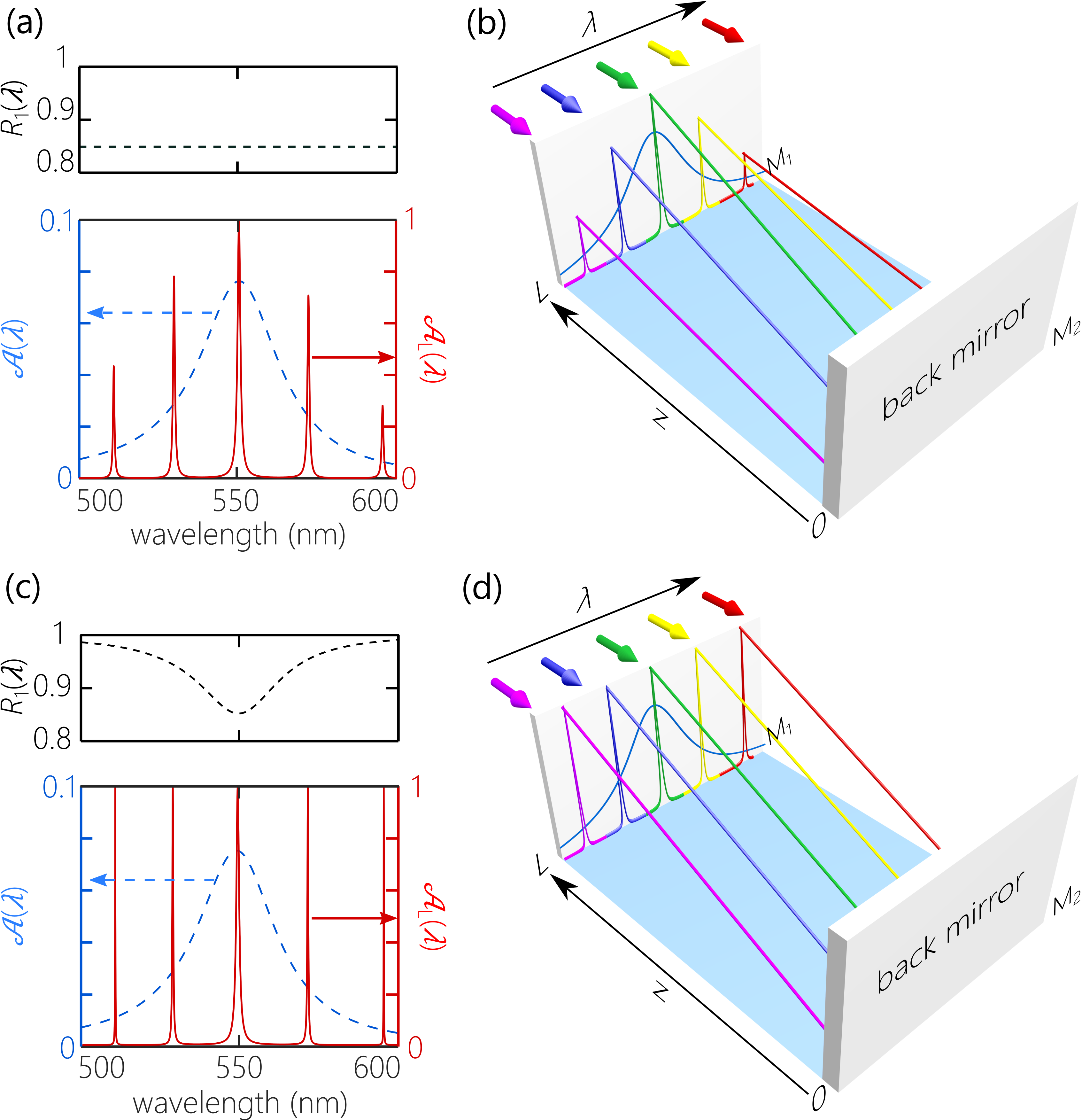}
\caption{\small{(a) Single-pass absorption $\mathcal{A}(\lambda)$ and total absorption $\mathcal{A}_{\mathrm{L}}(\lambda)$ for a cavity whose front mirror has a flat spectral reflectivity $R_{1}(\lambda)$; CPA is achieved only at the central wavelength. (b) Axial decay of the normalized Poynting's vector along the cavity at the resonant wavelengths. Except for the central wavelength, the field is not critically coupled to the cavity, which results in only partial absorption. (c) Same as (a) for a CPA cavity whose front mirror has an appropriately designed reflectivity $R_{1}(\lambda)$. CPA is achieved across the whole spectrum. (d) Same as (b), except that the field is critically coupled at all the resonances.}}
\label{Fig:Theory}
\end{figure}

We first elucidate our strategy for achieving broadband CPA by considering a thin layer (thickness $L$) of a material (refractive index $n$) having a single-pass intensity absorption $\mathcal{A}(\lambda)$ placed in a planar cavity with front and back mirrors M$_1$ and M$_2$ of reflectivities $R_{1}$ and $R_{2}$, respectively. A field incident normally on the cavity from the left produces transmitted $T_{\mathrm{L}}$ and reflected $R_{\mathrm{L}}$ waves, whereupon the total cavity absorption is $\mathcal{A}_{\mathrm{L}}$,
\begin{equation}
\mathcal{A}_\mathrm{L}(\lambda)\!=\!1-T_{\mathrm{L}}-R_{\mathrm{L}}\!=\!\frac{(1-R_1)(1-\Gamma)(1+\Gamma R_2)}{(1-\Gamma\!\sqrt{R_1R_2})^2+4\Gamma\!\sqrt{R_1R_2}\sin^{2}{\frac{\varphi}{2}}};
\end{equation}
where $\Gamma(\lambda)=1-\mathcal{A}(\lambda)$, $\varphi(\lambda)\!=\!2\frac{2\pi}{\lambda}nL\!+\!\alpha_{1}\!+\!\alpha_{2}$ is the round-trip phase, and $\alpha_{1}$ and $\alpha_{2}$ are the reflection phases from the M$_1$ and M$_2$, respectively. Realizing full absorption $\mathcal{A}_{\mathrm{L}}\!=\!1$ implies that $T_{\mathrm{L}}\!=\!R_{\mathrm{L}}\!=\!0$, which requires satisfying the following conditions: (i) $\varphi\!=\!m\pi$ for integer $m$, i.e., CPA occurs only at the cavity resonances; (ii) $R_2(\lambda)\!=\!1$, which implies the need for a back-reflector; and (iii) $R_{1}(\lambda)\!=\!(1-\mathcal{A}(\lambda))^{2}$, which necessitates utilizing a multilayer structure to produce a spectral reflectivity $R_{1}(\lambda)$ that tracks the material absorption. Three configurations are shown in Fig.~\ref{Fig:Concept}. First, spectrally flat $\mathcal{A}(\lambda)$ (e.g., graphene) requires utilizing a mirror with constant $R_{1}(\lambda)$ [Fig.~\ref{Fig:Concept}(a,b)]. Previous narrowband CPA observations fall into this class in which the wavelength-dependence of the intrinsic material absorption can be ignored. Second, monotonically \textit{decreasing} $\mathcal{A}(\lambda)$ (e.g., Si in the near-infrared) requires monotonically \textit{increasing} $R_{1}(\lambda)$ [Fig.~\ref{Fig:Concept}(c,d)], as demonstrated in \cite{Pye17OL}. Third, resonant absorption (e.g., in an organic laser dye) requires a spectral \textit{dip} in $R_{1}(\lambda)$ [Fig.~\ref{Fig:Concept}(e,f)]. This latter configuration has \textit{not} been realized to date to the best of our knowledge.

We consider a generic Lorentzian profile for the absorbing layer
$\mathcal{A}(\omega)\!=\!\mathcal{A}_{0}\frac{(\Delta\omega/2)^{2}}{(\omega-\omega_{0})^{2}+(\Delta\omega/2)^{2}}$, where $\omega_{0}$ is the resonant frequency, $\Delta\omega$ is FWHM of the Lorentzian profile, and $\mathcal{A}_{0}$ is its peak absorption [Fig.~\ref{Fig:Concept}(e)]. Placing such a layer in a planar cavity with with a back-reflector $R_{2}(\lambda)\!=\!1$ and a front mirror with flat spectral reflectivity $R_{1}$ can help realize CPA at one resonance (e.g., at $\lambda\!=\!\lambda_{0}$ if $R_{1}\!=\!(1-\mathcal{A}_0)^2$), but the absorption drops at all the others resonances [Fig.~\ref{Fig:Theory}(a)] because their coupling coefficients to the cavity (normalized with respect to the incident vales) are less than unity [Fig.~\ref{Fig:Theory}(b)].  Poynting's vector along the cavity in presence of the back-reflector is $P(z)\!=\!\mathcal{A}_{\mathrm{L}}(\lambda)\frac{\sinh{[\alpha(\lambda)z]}}{\sinh{[\alpha(\lambda)L]}}$, where $\alpha(\lambda)$ is the absorption coefficient, $\Gamma(\lambda)\!=\!e^{-\alpha(\lambda)L}$, and $z$ is measured from M$_2$ [Fig.~\ref{Fig:Theory}(b)]. In contrast, if the front mirror reflectivity is $R_{1}(\lambda)\!=\!(1-\mathcal{A}(\lambda))^{2}$, which thus has an appropriately shaped dip at $\lambda\!=\!\lambda_{0}\!=\!2\pi c/\omega_{0}$ ($c$ is speed of light in vacuum), then \textit{all} the resonances exhibit complete absorption $\mathcal{A}_{\mathrm{L}}\!=\!1$ across the whole spectrum independently of $\mathcal{A}(\lambda)$ [Fig.~\ref{Fig:Theory}(c)]. In this case, the field coupling to the cavity is unity at all wavelengths, and the normalized Poynting's vector decays according to $P(z)\!=\!\frac{\sinh{[\alpha(\lambda)z]}}{\sinh{[\alpha(\lambda)L]}}$ [Fig.~\ref{Fig:Theory}(d)]. In a weakly absorbing layer $\alpha(\lambda)\!\ll\!\frac{1}{L}$, Poynting's vector decays linearly along the cavity, $P(z)\!\approx\!\frac{z}{L}$, independently of $\alpha(\lambda)$ and thus also of $\lambda$. Moreover, perfect resonant absorption is possible even if the intrinsic absorption $\mathcal{A}(\lambda)$ does \textit{not} conform to a Lorentzian profile, as long as $R_{1}(\lambda)$ is judiciously engineered.

\begin{figure}[t!]
\centering\includegraphics[width=8.6cm]{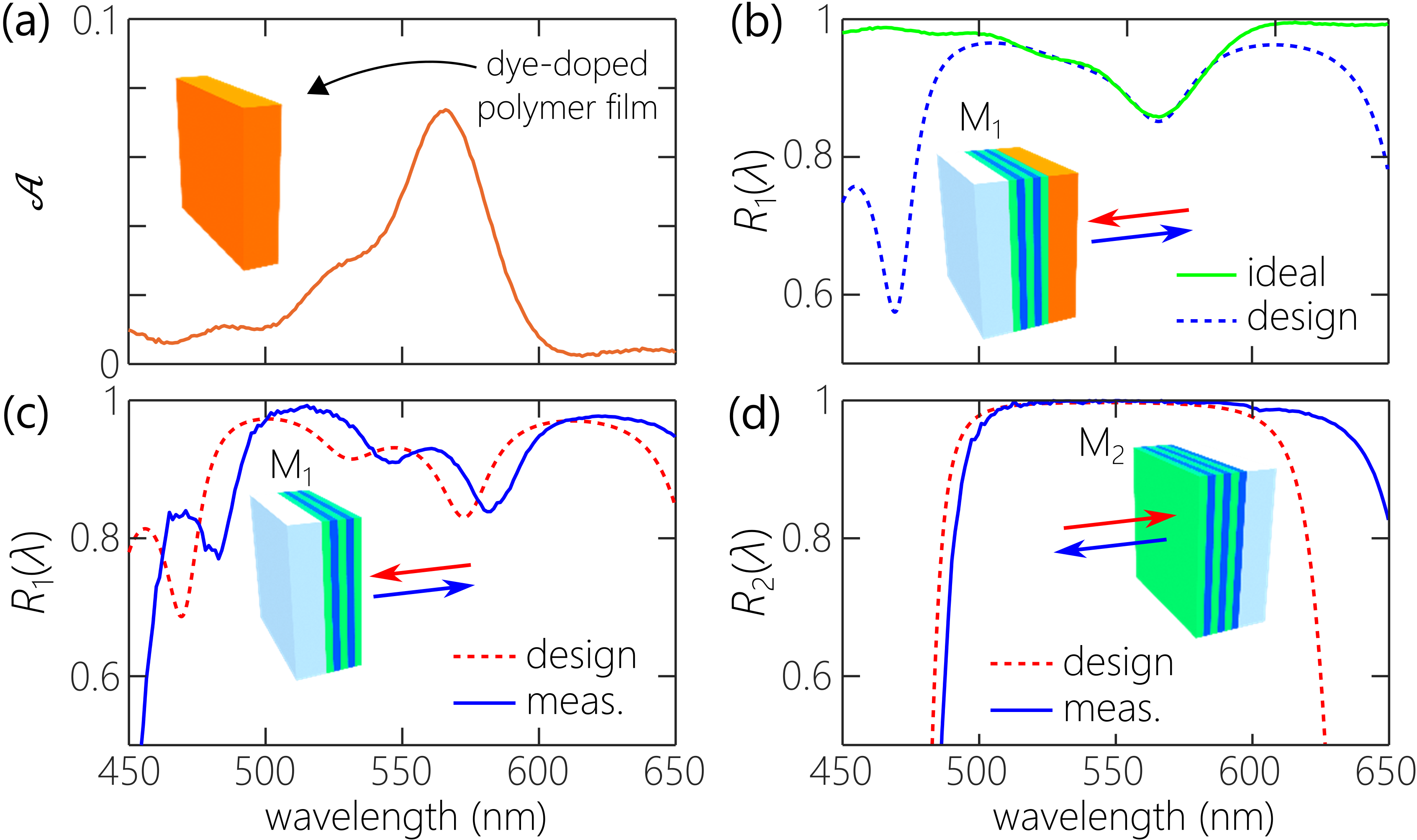}
\caption{\small{(a) Measured single-pass absorption $\mathcal{A}(\lambda)$ for a 2-$\mathrm{\mu m}$-thick PMMA layer doped with rhodamine 101. (b) The ideal reflectivity $R_{1}(\lambda)\!=\!(1-\mathcal{A})^{2}$ to realize CPA for the layer in (a) and the approximate design over the bandwidth of interest $500$-$600$~nm (assuming incidence from PMMA). (c) The designed and measured front mirror reflectivity $R_{1}(\lambda)$ for incidence from free-space. (d) The designed and measured back mirror reflection $R_{2}(\lambda)$.}}
\label{Fig:Mirror}
\end{figure}

Guided by this model, we realize CPA over a broad bandwidth ($\sim\!60$~nm) in a weakly absorbing 2-$\mathrm{\mu m}$-thick PMMA film doped with rhodamine 110, having a single-pass absorption peak of $\mathcal{A}_0\!\approx\!7.5\%$ [Fig.~\ref{Fig:Mirror}(a)]. We designed a multilayer front mirror M$_1$ (using the software package FilmStar) whose reflectivity approximates the ideal target of $R_{1}(\lambda)\!=\!(1-\mathcal{A}(\lambda))^{2}$ over the spectral range $500-600$~nm, and comprises 21~bilayers of SiO$_2$ (refractive index is $n\!\approx\!1.5$) and TiO$_2$ ($n\!\approx\!2.4$) [Fig.~\ref{Fig:Mirror}(b)]. We fabricated this front mirror by physical vapor deposition, and the measured reflection spectrum is shown in Fig.~\ref{Fig:Mirror}(c). The formula $R_{1}(\lambda)\!=\!(1-\mathcal{A}(\lambda))^{2}$ refers to the reflectivity of a field incident from the absorbing layer and \textit{not} from free space \cite{Villinger15OL}. The calculated spectrum in Fig.~\ref{Fig:Mirror}(c) takes into consideration the necessary modification in $R_{1}(\lambda)$ \cite{Pye17OL}. The measurement reveals an excellent match with theory except for a $\approx\!\!10~\mathrm{nm}$ spectral red-shift with respect to the ideal spectrum. The back-reflector M$_2$ is a periodic Bragg reflector comprising 21 alternating quarter-wavelength layers of SiO$_2$ and TiO$_2$, which yields a near-unity flat spectral reflectivity $R_{2}(\lambda)\!\approx\!1$ [Fig.~\ref{Fig:Mirror}(d)].

\begin{figure}[t!]
\centering\includegraphics[width=8.6cm]{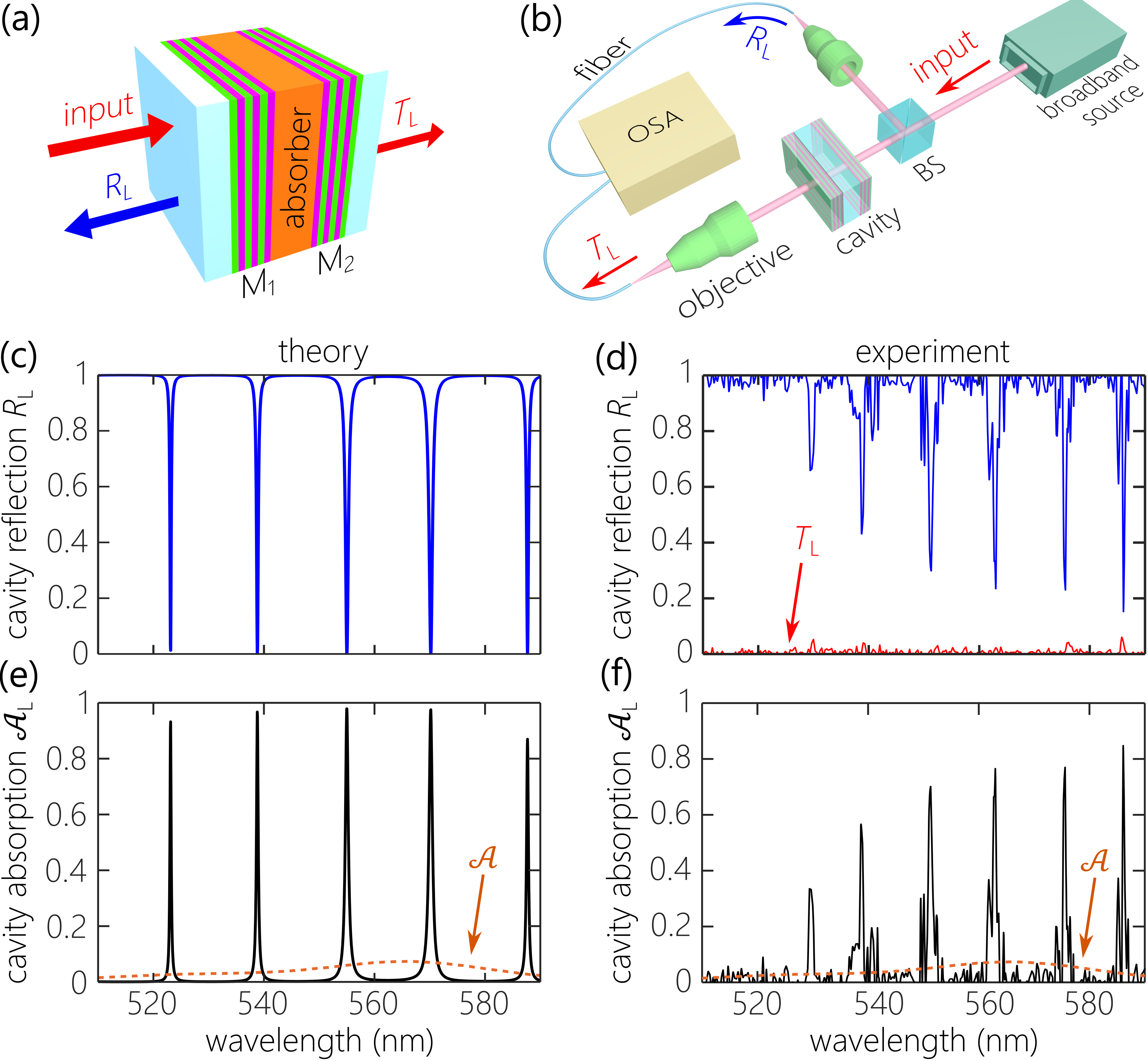}
\caption{\small{(a) CPA cavity structure. (b) schematic of the setup for to measure $\mathcal{A}_{\mathrm{L}}(\lambda)$. The transmitted and reflected fields are collected and coupled via fibers to an optical spectrum analyzer (OSA). (c) The calculated and (d) measured cavity reflection $R_{\mathrm{L}}(\lambda)$; the transmission $T_{\mathrm{L}}(\lambda)$ is negligible because $R_{2}\!\approx\!1$. (e) The calculated and (f) measured the cavity absorption $\mathcal{A}_{\mathrm{L}}(\lambda)\!=\!1\!-\!T_\mathrm{L}(\lambda)\!-\!R_\mathrm{L}(\lambda)$. A large enhancement in the resonant absorption is clear compared to the single-pass absorption $\mathcal{A}(\lambda)$.}}
\label{Fig:RhodamineCavity}
\end{figure}

Using the mirrors M$_1$ and M$_2$ whose reflectivities satisfy the desiderata for CPA, we construct an asymmetric planar cavity by spin-coating a PMMA-dye solution on the back mirror and then sandwiching the polymer layer between the mirrors [Fig.~\ref{Fig:RhodamineCavity}(a)]. We measured the cavity absorption $\mathcal{A}_{\mathrm{L}}(\lambda)$ utilizing the setup depicted schematically in Fig.~\ref{Fig:RhodamineCavity}(b). A collimated beam from a broadband incoherent light source (a Tungsten lamp) is directed to the cavity and the transmitted $T_{\mathrm{L}}$ and reflected $R_{\mathrm{L}}$ beams are collected and delivered via multimode fibers (50-$\mathrm{\mu m}$ core diameter) to an optical spectrum analyzer (OSA), from which we obtain the total absorption $\mathcal{A}_{\mathrm{L}}\!=\!1\!-\!T_{\mathrm{L}}\!-\!R_{\mathrm{L}}$. We plot in Fig.~\ref{Fig:RhodamineCavity}(c,d) the calculated and measured spectral reflectivity $R_{\mathrm{L}}(\lambda)$; the measured spectral transmission $T_{\mathrm{L}}(\lambda)$ is negligible as expected. The calculated and measured spectral absorption in the cavity $\mathcal{A}_{\mathrm{L}}(\lambda)$ are plotted in Fig.~\ref{Fig:RhodamineCavity}(e,f), along with the measured single-pass absorption $\mathcal{A}(\lambda)$ from Fig.~\ref{Fig:Mirror}(a). 

We can draw several conclusions from the results in Fig.~\ref{Fig:RhodamineCavity}. First, the structure produces an absorption enhancement $\mathcal{A}_{\mathrm{L}}(\lambda)/\mathcal{A}_{0}\!\approx\!12-25$ at the cavity resonances. Second, this absorption enhancement is produced over a broad bandwidth $\sim\!60$~nm, which extends to the edges of the dye absorption linewidth. Even resonances that are strongly displaced from the rhodamine absorption-peak wavelength exhibit significantly higher absorption $\mathcal{A}_{\mathrm{L}}(\lambda)$ than the single-pass absorption $\mathcal{A}(\lambda)$ at the same wavelength. The mirror design guarantees a perfect resonant absorption across the whole bandwidth, independently of the intrinsic spectral absorption of the organic film. The decay in the absorption enhancement at the edges of this spectral range is likely due to the deviation in $R_{1}(\lambda)$ of the fabricated front mirror with respect to the targeted design [Fig.~\ref{Fig:Mirror}(c)].

\begin{figure}[t!]
\centering\includegraphics[width=8.6cm]{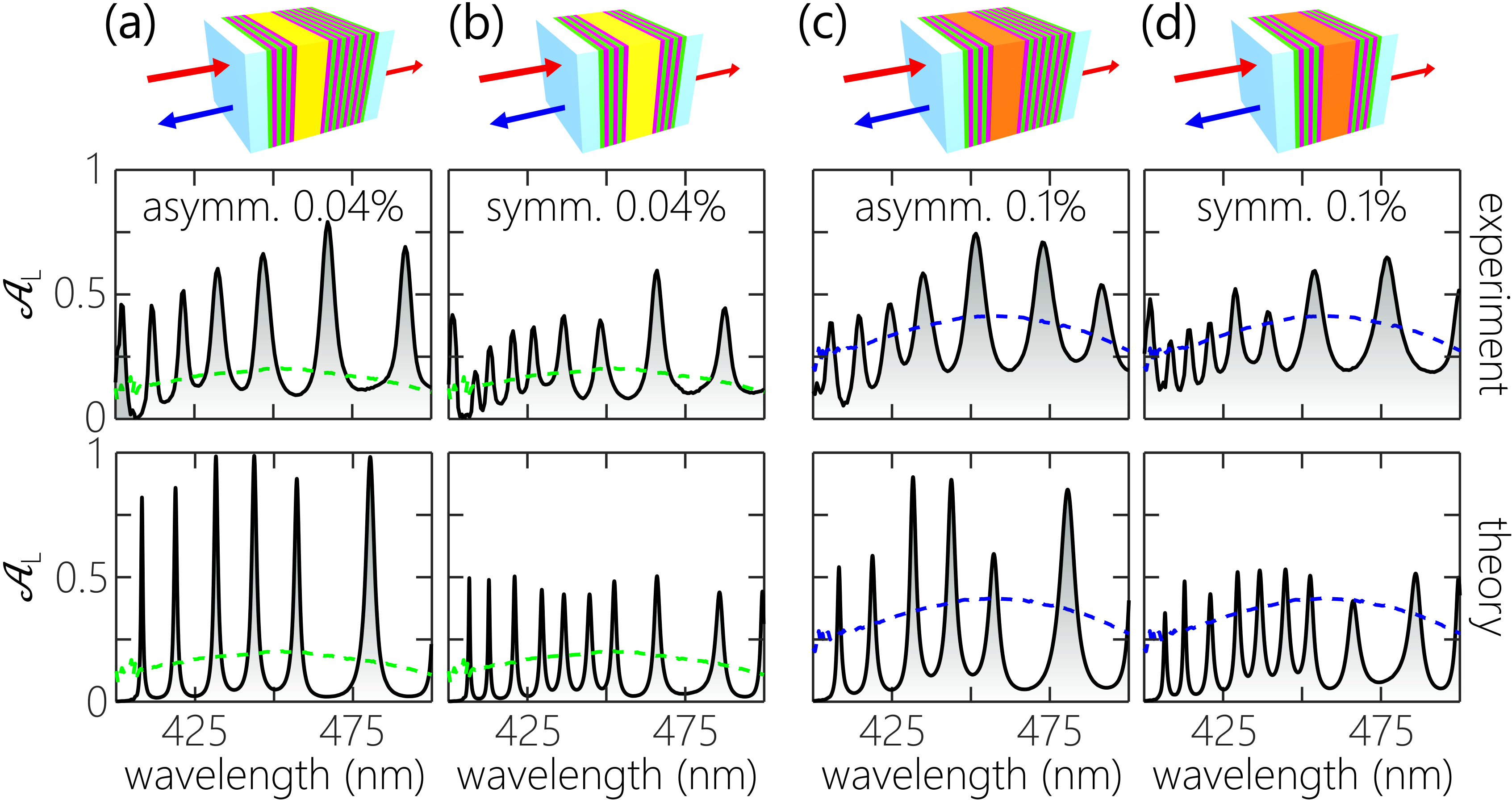}
\caption{\small{(a) Absorption spectrum for two $\approx\!2$-$\mathrm{\mu m}$-thick films of DCM-doped PMMA at different concentration levels. (b) The symmetric (right column) and asymmetric (left column) cavity for a DCM concentration of $0.04\%$. The cavities are illuminated from left, and the measured absorption is compared with theoretical expectations (bottom row). The results indicate the enhanced resonant absorption can reach close to 100\% in the asymmetric device,  while it cannot increase much further beyond $50\%$ for the symmetric cavity. (c) Same as (b), with $0.1\%$ DCM concentration. The dashed curves throughout are the single-pass absorption $\mathcal{A}(\lambda)$.}}
\label{Fig:DCMSymmetry}
\end{figure}

To further confirm this effect, we implemented the same procedure with DCM. We tested two single-pass absorption levels $\mathcal{A}$ produced by varying the DCM concentration in a $\approx\!2$-$\mathrm{\mu m}$-thick PMMA layer [Fig.~\ref{Fig:DCMSymmetry}]. Furthermore, we studied the absorption in two different cavity structures: (1) an \textit{asymmetric} structure similar to that utilized above with $R_{1}(\lambda)\!=\!(1-\mathcal{A}(\lambda))^{2}$ and $R_{2}(\lambda)\!=\!1$, whereupon CPA ($\mathcal{A}_{\mathrm{L}}\!=\!1$) can be ideally achieved; and (2) a \textit{symmetric} cavity in which $R_{1}(\lambda)\!=\!R_{2}(\lambda)$. In the latter case, the cavity achieves an \textit{optimal} absorption of $\mathcal{A}_\mathrm{L}\!=\!\frac{1}{8}\frac{(2-\mathcal{A})^2}{1-\mathcal{A}}$ when $R_{1}(\lambda)\!=\!R_{2}(\lambda)\!=\!1\!-\!\mathcal{A}(\lambda)$ \cite{Villinger15OL}. Here CPA cannot be achieved. Instead, an optimal absorption of \textit{at least} $\mathcal{A}_{\mathrm{L}}\!=\!\frac{1}{2}$ is guaranteed even for vanishingly small $\mathcal{A}$, and a maximum optimal absorption of $\mathcal{A}_{\mathrm{L}}\!=\!\frac{2}{3}$ is reached when $\mathcal{A}\!=\!\frac{2}{3}$ (for $\mathcal{A}\!>\!\frac{2}{3}$, the optical absorption is $\mathcal{A}_{\mathrm{L}}\!=\!\mathcal{A}$) \cite{Villinger15OL}.

\begin{figure}[t!]
\centering\includegraphics[width=8.6cm]{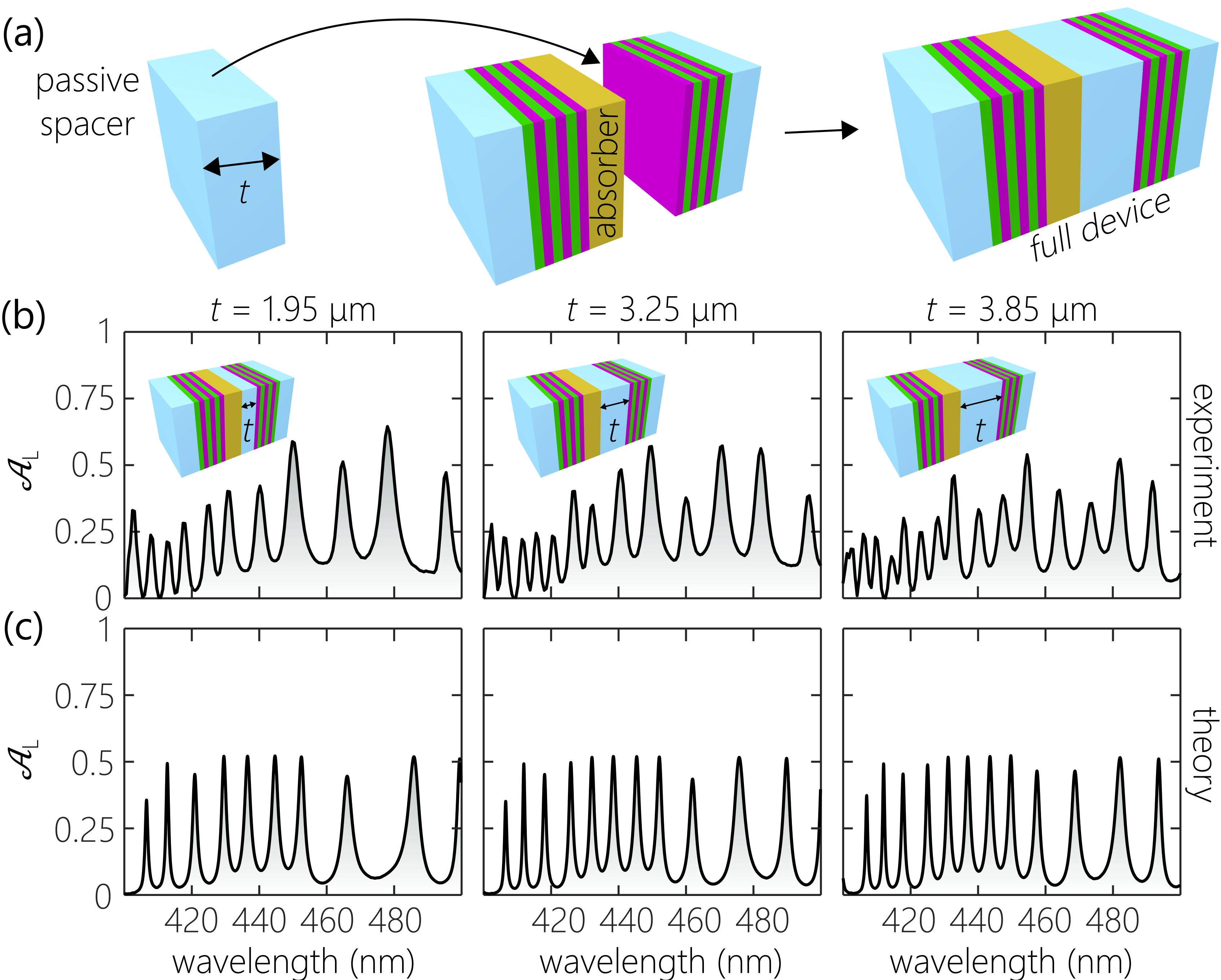}
\caption{\small{(a) Construction of a symmetric planar cavity containing a $\approx\!2$-$\mathrm{\mu m}$-thick polymer film doped with $0.1\%$ DCM dye [Fig.~\ref{Fig:DCMSymmetry}(c,d)] and also incorporates a transparent silica spacer. Three such devices are produced with different spacer thickness $t$ adjacent to the absorbing film. (b) The measured and (c) the calculated absorption spectra for spacer thicknesses of $t\!\!=1.95 \mathrm{\mu m}$ (left column), $t\!\!=3.25 \mathrm{\mu m}$ (middle column) and $t\!\!=3.85 \mathrm{\mu m}$ (right column). The results indicate no significant impact of the extra spacer thickness on the resonantly enhanced absorption, while it increases the number of resonances (reduces the FSR) because of the increased cavity length.}}
\label{Fig:DCMspacer}
\end{figure}

Calculated and measured absorption spectra are plotted in Fig.~\ref{Fig:DCMSymmetry}(a,b) for the asymmetric and symmetric cavities for a DCM concentration of $0.04\%$, and for a DCM concentration of $0.1\%$ in Fig.~\ref{Fig:DCMSymmetry}(c,d). The mirrors in all four cavities are constructed of alternating layers of Nb$_2$O$_5$ ($n\!\approx\!2.25$) and SiO$_2$, optimized using the same approach outlined above. The results confirm both the broadband absorption enhancement in the cavity above that of the single-pass absorption, and also confirms that the absorption peaks for the symmetric cavities only slightly exceed $50\%$. Note that the absorption enhancement is larger for the case of lower dye concentration. 

Finally, we carried out an experiment to ensure that increasing the cavity size (thus reducing its FSR) by inserting a passive spacer within the cavity does not reduce the resonant enhancement in absorption, and can indeed be useful in engineering the absorption spectrum. Using the $0.1\%$-concertation DCM layer from Fig.~\ref{Fig:DCMSymmetry}(c,d), we construct three symmetric cavities using the same mirror reflectivites, but with different-thickness silica spacers deposited on M$_2$ [Fig.~\ref{Fig:DCMspacer}(a)]. The calculated and measured absorption spectra $\mathcal{A}_{\mathrm{L}}(\lambda)$ plotted in Fig.~\ref{Fig:DCMspacer}(b,c) confirm that a passive spacer does \textit{not} have a detrimental impact on the absorption except for increasing the number of resonances and reducing the FSR.

Our strategy to realize CPA is materials-agnostic, and is thus applicable to other material systems with different underlying absorption mechanisms. The measurements confirm CPA at the discrete wavelengths associated with the cavity structure. We have recently shown that these resonances can be broadened to extend over extended spectral ranges \textit{without} modifying the cavity itself via a strategy that we have called `omni-resonance' \cite{Shabahang17SR,Shabahang19OL,Shiri20OL}. Applying this approach here would yield broadband, continuous wavelength CPA in resonant optical materials. Finally, the approach outlined here when combined with omni-resonance can also benefit pump gain media \cite{Jahromi17ACSP} and even contribute to reduced-threshold lasing \cite{Jahromi17NC} and non-Hermitian optics in general.

In conclusion, we have proposed a generic design for achieving CPA in resonant materials such as organic laser dyes, and experimentally demonstrated the concept with rhodamine~101 and DCM dyes. An appropriate planar photonic environment containing a mirror featuring a spectral reflectivity dip guarantees that a weakly absorbing layer absorbs light fully, independently of its intrinsic absorption level -- realized with a flat absorption spectrum. Our materials-agnostic approach is independent of the spectral range or target bandwidth. This strategy that combines the simplicity of a planar structure with the convenience of a non-interferometric configuration (only a single incident beam) can be useful in a variety of applications, including photodetectors, solar cells, and optical switches.

\section*{Funding}
Air Force Office of Scientific Research (AFOSR) under MURI award FA9550-14-1-0037.

\section*{Acknowledgments} The authors are grateful to Nathan Bodnar for technical assistance.

\section*{References}

\bibliography{Refs}{}
\bibliographystyle{unsrt}
\end{document}